# Rethinking the Role of Technology in Virtual Teams in Light of COVID-19

**Short Paper**

**Mark Frost**
School of Management and Marketing
Charles Sturt University
Bathurst, Australia
Email: mfrost@csu.edu.au

**Sophia Xiaoxia Duan**
School of Accounting, Information Systems, and Supply Chain
RMIT University
Melbourne, Victoria
Email: sophia.duan2@rmit.edu.au

## Abstract

The use of virtual teams by organisations has grown tremendously as a strategic response to COVID-19. However, the concept of virtual teams is not something new, with many businesses over the past three decades gradually incorporating virtual and/or dispersed teams into their processes. Research on virtual teams has followed that of co-located face-to-face teams through lenses such as trust, communication, teamwork, leadership and collaboration. This paper introduces a new paradigm for examining the development of virtual teams, arguably one that would facilitate the consideration of technology as part of a virtual team rather than simply as an alternate to face-to-face teams. That is, viewing the development of virtual teams with embedded technology within an organisation through an innovation framework.

**Keywords**　　virtual teams, shared leadership, innovation, virtual organisations

# 1　Introduction

> In today's environment, one may never or seldom be physically co-located with other members of a work unit but be part of a virtual work unit, often using sophisticated IT systems such as cloud-based computing to coordinate and share projects, duties, and work-related information. For example, in telecare Nickelsen and Elkjaer (2017) note that in the ICT infrastructure dominated context of providing medical telecare, the ICT infrastructure shapes the work unit community and interactions; thereby, challenging the hierarchy and relationships of the work unit. (Arnold et al., 2019: 368)

One of the many impacts of COVID-19 has seen many organisations move their workforce to working from home and the increasing use of technology. This has seen increased commentary from publications such as the Economist, Guardian, UK Financial Times and the New York Times (to name a few) which suggest the end of the traditional "office" through the adoption of hybrid type work arrangements, and the benefits of utilising technology and the creation of virtual teams.

These changes are not new and have been occurring incrementally since the wider introduction of computers and other technology into the workplace since the 1990s. For example, Hoch and Dulebohn (2017) quote a 2014 RW3 survey of 3000 managers that 40% of their employees spend time working in virtual teams. A 2012 survey of 379 HR professionals by the Society of Human Resource Management finds that 46% of the participants reported their organizations to use virtual teams. Liao (2017) also refers to this latter survey as well as suggesting that co-located teams have increasingly incorporated technologies into work operations and to facilitate communication. Along similar lines, Cogliser et al. (2012) conclude that teams have become more dispersed over both time and space and utilising technology to facilitate this in varying degrees. In earlier research Siebdrat et al. (2009) conclude that such dispersion is not necessarily detrimental but depends on team's processes including team member contribution and coordination.

Research on virtual teams has followed that of co-located face-to-face teams through lenses such as trust, communication, teamwork, leadership and collaboration. In a report on developments in China business in response to COVID-19 the McKinsey and Company (2020) suggest that many teams have simply moved from face-to-face and co-located to virtual teams. In the majority of cases, these existing teams have simply recreated their existing work processes through increased utilisation of virtual technologies. While not suggested in the report this does raise the issue of the development of new teams or even the re-purpose of existing teams in the new virtual environment.

This short paper introduces a new paradigm for examining the development of virtual teams, arguably one that would facilitate the consideration of technology as part of virtual teams rather than simply as an alternate to the face-to-face teams. That is, viewing the development of virtual teams with embedded technology within an organisation through an innovation framework.

The remainder of this paper is organized as follows. Section 2 introduces related components of a good virtual team. Section 3 presents theories from the innovation and entrepreneurship viewpoints that are used to guide the discussion in Section 4. Section 5 raises some questions for further discussion. The last section draws the conclusion.

# 2　What makes a good virtual team?

Virtual teams refer to the geographically distributed collaborations that rely on technology to communicate and cooperate (Morrison-Smith & Ruiz, 2020). Existing literature on virtual teams is comprehensive (Gilson et al., 2014; Morrison-Smith & Ruiz, 2020) and beyond the scope of this short paper. The following section seeks to introduce some of the key concepts to generate discussion rather than attempting to provide an extensive review of literature. Many of the concepts raised apply to both virtual, dispersed and face-to-face teams. Virtual teams have an added layer of complexity related to the use of technology within the team. The literature below focuses largely on the human behaviour elements of virtual teams, with the role of technology introduced later.

Hoch and Dulebohn (2017) suggest that trust is an important component of any team. Shared leadership and collaboration between members increase trust and knowledge sharing, improve functioning and assist with the achievement of individual and unit goals. Shared leadership is considered a mutual influence process to assist the achievement of objectives. However, the authors also conclude that co-located teams work more effectively as a lack of co-located interaction impacted trust, decision-making, conflict management and the expression of opinions. This is consistent with the findings of Siebdrat et al. (2009) above.

The human dimension is also important to the success of virtual teams. Such a dimension maximises collegiality, engagement and civility (Arnold et al., 2019). For virtual teams, the composition of member personality is important as is conscientiousness, agreeableness and emotional stability (Hoch and Dulebohn, 2017). Conscientiousness implies a strong sense of direction, self-discipline and good self-organisation skills. Agreeableness is linked with being good-natured and cooperative. Emotional stability suggests a calm, self-confident team member that possesses high self-efficacy, and could also include tolerance of both ambiguity and uncertainty. Siebdrat et al. (2009) suggest that when selecting members, it is important to consider member social skills and the underlying self-sufficiency. This is important for all dispersed and virtual teams as the development of underlying group dynamics is not as extensive as what usually occurs in a face-to-face or co-located setting.

Emotional intelligence is a significant predictor of unit effectiveness in face-to-face teams. However, virtual teams observe fewer see verbal and nonverbal clues due to decreased social interaction and emotional expression. Emotional intelligence is a critical driver in successful outcomes within a virtual team. Similarly, this lack of social interaction sees team development occurring more slowly in virtual teams (Pitts et al., 2012).

Gressgard (2011) concludes that highly functioning virtual teams facilitate flexible approaches to product development, shortens innovation processes, which in turn speeds up the time to market, thereby providing a competitive advantage. Furthermore, digital disruption now sees a merging of information systems and processes such that production, distribution, information and communication systems can now be linked together.

Leung et al. (2020) suggest that an emerging response from the COVD-19 is hybrid working arrangements, i.e. spilt time between the workplace and home. This may see more discussion on work/life integration as distinct from work/life balance. It will also mean that both existing and new virtual teams may need to repurpose work processes rather than recreate existing processes in the hybrid arrangements. This is consistent with Larson and DeChurch (2020) who suggest that we are now entering a new phase of virtual teams. They suggest that there are four themes linking technology, virtual teams and organisational processes. These are

(i) Technology as context – where technology is a fixed feature that sets the context for unit processes, communication and information storage. An example is seen in the 1990s with the widespread introduction of computers in organisations and the 2000s with the evolution and growth of virtual teams. Figure 1 highlights the evolving use of technology since the 1990s.

(ii) Technology as socio-material – technology and units are mutually dependent. This is more for sociological considerations that are outside the scope of this short paper.

(iii) Technology as a creation medium – where units are formed through technology such as flash teams, project groups relating to the introduction of new automated processes or organisational systems, but also sometimes outside of formal structures like Wikipedia.

(iv) Technology as a teammate – this is an emerging theme and occurs where technology has a distinct role as an effective virtual team member or the core that the virtual team is built around. In the above three themes, the virtual teams work with technology through either communication or process. In this theme, technology has a distinct role as a distinct member and may include the use of robotics, algorithms or artificial intelligence. The authors suggest that this may be the next major development in the use of technology.

To date, the majority of technological improvement has allowed incremental change within organisations, however disruptive and transformational change can also provide an opportunity for organisations (Gressgard, 2011). Figure 1 shows the advancement of technologies in supporting virtual teams.

In short, the notion of virtual teams is not something new that has simply emerged from COVID-19. Rather there has been ongoing development of virtual and dispersed teams for nearly three decades, with accompanying research on the performance, leadership and outcomes. Attributes of successful virtual teams include shared leadership, emotional stability/intelligence, conscientiousness, and agreeableness. Arguably most of this development has occurred in the theme of technology providing a context. Much of the literature on virtual teams have utilised existing face-to-face team literature as a framework, where teams have been re-creating existing processes and structures rather than as a tool for re-purposing.

Utilising the theme of technology as a teammate may provide an opportunity for transformational change within an organisation as it emerges from the impacts of COVID-19. It would see emerging technologies such as robotics, automation and/or artificial intelligence be considered concurrently when team development occurs and how team members could be built these technologies to maximise opportunities. The question is how could organisations enable this?

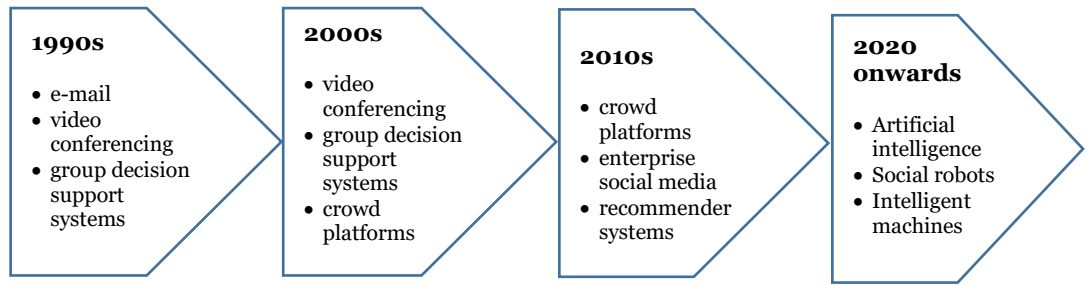

**Figure 1.**  Technology advancement in supporting virtual work units

## 3 Entrepreneurship and Innovation

To facilitate the development of a framework for understating the new paradigm of virtual teams post COVID-19, the body of literature on entrepreneurship and innovation is reviewed. Existing literature from these two streams has different viewpoints on virtual teams (Gilson et al., 2014; Larson and DeChurch, 2020; Morrison-Smith & Ruiz, 2020).

The entrepreneurship literature is extensive and this short paper mainly investigates the organisational process for facilitating the virtual team. Some seminal works on entrepreneurship and innovation are introduced in the following commentary. Entrepreneurship can be considered a process of "creative destruction", where the entrepreneur displaces or destroys the existing methods of production and replaces these with new ones, be it in new products or services, new methods of production, new markets, new supply chains or new forms of production (Schumpeter 1934). Burgelman (1983) defines corporate entrepreneurship as the process where firms (as distinct from individuals) engage in diversification through new resource combinations that extend the firm's activities into unrelated areas, or marginally related to its current operations. Burgelman and Doz (2011) subsequently assed that firms need both order and diversity in their strategy to survive. Order is provided in management and structural processes while corporate entrepreneurship provides a diversity through experimentation and selection. Covin and Slevin (1989) find that in hostile environments a more adaptive entrepreneurial strategic posture can be useful. Covin and Miles (1999) coin the term organisational rejuvenation as when an organisation alters its exiting processes, structures or capabilities in face of such hostile environments. The authors identified that innovation underlies all forms of corporate entrepreneurship, either through a product, service, or part of the wider organisational direction (Covin & Miles, 1999).

The innovation literature primarily examines the adoption and diffusion of virtual teams in organisations and the factors associated with the adoption and diffusion. Prominent theoretical frameworks for investigating the adoption and diffusion of technologies in virtual teams include the theory of reasoned action, the theory of planned behaviour, the diffusion of innovation theory, the technology acceptance model, the unified theory of acceptance and use of technology, the motivational model theory, and the technology-organisation-environment framework (Guth and Ginberg, 1990; Zahra 1993; Covin and Miles, 1999: 2007; Venkatesh, et al., 2016; Kahn, 2018). The innovation in virtual teams can be seen as an outcome from a goal (i.e. in process, output, or product, etc.); a mindset (i.e. innovation at an individual level, a supportive team structure and /or culture); or a process (how does it affect current processes) (Kahn, 2018). Therefore innovation can be internally focused and include cost reduction, product improvement, new markets, efficiency in processes, amended supply chains to name a few. While not stated perhaps technology could also be included in these items as it links easily with all of these processes (Covin and Miles, 1999: 2007).

Both the entrepreneurship literature and the innovation literature provide rich information in understanding different aspects of virtual teams. Considering the new role of technology being the teammate in virtual teams, the overlap of the two streams of research sheds light on the key theories and factors for constructing a framework for understating the new paradigm of virtual teams post COVID-19. The next section will cover the initial discussion on constructing such a framework.

## 4  Discussion

As businesses emerge from their amended operational arrangements from COVID-19 they will face a changing landscape to how they have operated pre COVID-19. The implementation of virtual teams, while widely reported in media as a relatively new paradigm, has been incrementally occurring since the 1990s. In many cases, existing teams have simply recreated existing processes to allow for amended working practices, both within the organisation and also with external stakeholders. It is unlikely that businesses will simply return to where they were pre COVID-19. For example, teams may retain an element of dispersion and/or virtual components. The work-life integration of team members may become more prevalent such that team members are working outside existing business hours in the form of hybrid working arrangement.

There may be an opportunity to repurpose existing teams as well as develop new teams that also include further adoption of technological features such that technology is considered an effective "teammate". This can include the use of existing tools for sharing resources and communication as well as the increased potential use of robotics, automation and artificial intelligence. The creation of these teams may require a different mindset around staff recruitment, performance and development, management structures and ways to measure success. There is a general understanding that formal leadership in dispersed and virtual teams require more focus on group and individual communication as well as a focus of group outputs.

Recruitment may need to be adjusted to also include technological understanding, emotional intelligence and self-efficacy. Furthermore, formal leaders may need to ensure strong communication channels exist to develop team functioning such that trust and shared leadership emerge.

The characteristics for successful virtual team members include factors such as shared leadership, emotional stability/intelligence, conscientiousness, and agreeableness. The majority of these attributes match those of a successful entrepreneur, albeit some more directly than others. Entrepreneurial attributes include independence, self-efficacy, a propensity for risk and opportunity, a drive to achieve, a creative thinker, a problem solver with a high tolerance for ambiguity and/or failure, and have team-building skills. All of these attributes would be well suited within a virtual team environment. Figure 2 shows a framework for formulating a high performing virtual team post COVID-19.

This suggests that organisations looking to repurpose existing processes post COVID-19 with possible more detailed engagement with technology could do so within an entrepreneurial framework. Namely flexible and entrepreneurial organisational structures, a goal approach to success, and an acute awareness of industry trends. Interestingly the two seminal pieces on entrepreneurship literature occurred in response to significant economic upheavals. This is probably understandable as it links with the commentary on succeeding in hostile environments. A key question is whether current circumstances could be considered something familiar.

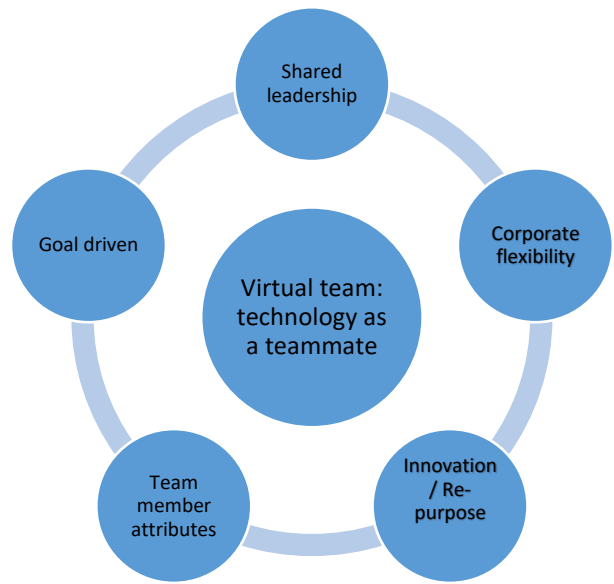

**Figure 2.**　　　An innovation framework for virtual team

From Figure 2 a few questions arise, namely:

- Is this an opportunity to re-purpose existing processes and organisational structures?
- If so, what is the role of technologies in virtual teams post COVID-19?
- What will be the success factors for the high performing virtual teams post COVID-19?
- What are the behaviours of high performing virtual teams post COVID-19?
- How could organisations enable and empower the virtual team post COVID-19?
- Will technology have a greater role than what currently exists?

## 5 Conclusion

This short paper introduces a new paradigm for examining the development of virtual teams post COVID-19 with the consideration of technology as part of virtual teams. That is, rather than being an enabler of process, can technology (and how it is utilised) influence the dynamics of a team, all people and process activities related to the team, and the ability of the team to innovate more successfully. Possible linkages between the development of successful virtual teams, and how the attributes of these teams could assist corporate innovation and vice versa are discussed. To inspire further discussion on the new paradigm of virtual teams, open questions are proposed in line with the goals of this paper. Namely, the introduction of a possible framework that businesses could utilise as they grapple with the amended demands of their environment post COVID-19.